\documentclass[12pt]{article}

\usepackage[english]{babel}

\usepackage[letterpaper,top=2cm,bottom=2cm,left=3cm,right=3cm,marginparwidth=1.75cm]{geometry}

\usepackage{amsmath}
\usepackage{graphicx}
\usepackage[colorlinks=true, allcolors=blue]{hyperref}

\usepackage{multirow}
\usepackage{tabularx}
\usepackage{subcaption}
\usepackage{comment}
\usepackage{bm}
\usepackage{algorithm2e}
\usepackage{authblk}

\usepackage{float}

\newcommand{\absdiv}[1]{%
  \par\addvspace{.5\baselineskip}
  \noindent\textbf{#1}\quad\ignorespaces
}

\newcommand{\killpunct}[1]{}

\usepackage[square,numbers,sort]{natbib}
\usepackage{doi}

\title{Colorectal cancer risk mapping through Bayesian networks}

\author[a]{D. Corrales\thanks{Corresponding author. Email: daniel.corrales@icmat.es}}
\author[b,c]{A. Santos-Lozano}
\author[c]{S. López-Ortiz}
\author[b,d]{A. Lucia}
\author[a]{D. Ríos Insua}

\affil[a]{Inst. Math. Sciences, CSIC, 28049 Madrid, Spain}
\affil[b]{Research Institute of Hospital 12 de Octubre (‘imas12’), 28041 Madrid, Spain}
\affil[c]{i+HeALTH Strategic Research Group, Miguel de Cervantes European University, 47012 Valladolid, Spain}
\affil[d]{Faculty of Sport Sciences, Universidad Europea de Madrid}

\date{  }

\begin{document}

\maketitle
\begin{abstract}
\absdiv{Background and Objective} \textcolor{black}{ Only about 14 \% of eligible EU citizens finally participate} in colorectal cancer (CRC) screening programs despite \textcolor{black}{it} being the third most common type of cancer worldwide. The development of \textcolor{black}{ CRC risk models can enable predictions to be embedded in decision-support tools facilitating CRC screening and treatment recommendations. This paper develops a predictive model that aids in characterizing CRC} risk groups and assessing the influence of a variety of risk factors on the population.

\absdiv{Methods} A CRC Bayesian Network is learnt by aggregating extensive expert knowledge and data from an observational study and making use of \textcolor{black}{ structure learning algorithms to 
 model the relations between variables}. The network is then parametrized to characterize these relations in terms of local probability distributions at each of the nodes. It is finally used to predict the \textcolor{black}{ risks of developing CRC together with the uncertainty around such predictions}.

\absdiv{Results} A \textcolor{black}{ graphical CRC risk mapping tool } is developed from the model and used to segment the population into risk subgroups according to variables of interest. Furthermore, the network provides insights on the \textcolor{black}{predictive} influence of modifiable risk factors such as alcohol consumption and smoking, and medical conditions such as diabetes or hypertension linked to \textcolor{black}{ lifestyles that   potentially have an impact on an increased risk of}
 developing CRC.

\absdiv{Conclusions} CRC is most commonly developed in older individuals. \textcolor{black}{However, some modifiable behavioral factors seem to have a strong predictive influence on its potential risk of development}. Modelling these effects facilitates identifying risk groups and targeting influential variables which are subsequently helpful in the design of screening and treatment programs.

\absdiv{Keywords} Colorectal cancer, Bayesian network, Risk mapping, Modifiable risk factors, Health policy. 
\end{abstract}

\section{Introduction}

Colorectal cancer (CRC) is the third most common type of cancer worldwide, 
making up for about 10\% of all cases \cite{WHO} and being accountable for
around 12\% of all deaths due to cancer. In 2020, there were 1.9 million new cases and 930,000 associated deaths. It is more common in developed countries, where more than 65\% of the cases are found. 
Despite this, as an example, only about 14\% of susceptible EU citizens participate in screening programs, 
 at the moment mostly based on \textcolor{black}{ fecal testing and colonoscopy. 
 Hence, there is a need for} accurate, non-invasive, cost-effective screening tests based on novel technologies and raise further awareness of the disease and its detection. 
  Moreover, more personalized screening approaches are required to consider genetic and socioeconomic variables as well as environmental stressors that \textcolor{black}{ potentially lead } to different onsets of the disease \citep{KASTRINOS}. 
A particular line of action is the development of predictive models that facilitate CRC predictions, the subject of this paper, possibly embedded in decision support tools \textcolor{black}{that aid in the} advice on screening and treatment recommendations. 

The epidemiology of CRC and its most important risk factors (CRCRF) 
are discussed, among others,
in
\citet{Marley2016-hl} and \citet{sawicki2021}.
 These factors are defined as measurable characteristics associated with increased CRC incidence and considered to be significant independent predictors of increased risk of the disease.
   \textcolor{black}{ They are qualified as modifiable or not.} 
 Non-modifiable ones are factors over which the individual has no control, including genetics, age, or gender. In contrast, modifiable ones cover behavioral factors that can \textcolor{black}{evolve} through individual action, including physical activity (PA), or \textcolor{black}{tobacco use}. Most CRC development does not have a genetic burden, but is linked to lifestyle and environmental factors \cite{sawicki2021} and thus the identification of the impact of the modifiable factors in individuals is key to \textcolor{black}{reducing} CRC incidence.

The purpose of this paper is to provide a Bayesian network (BN)
\citep{Jensen1996, scutari2021bayesian} that facilitates the prediction of CRC risks and their mapping. The network will be built 
from extensive expert judgment and data and illustrated 
through two relevant use cases referring to CRC risk mapping and CRC influential finding identification;
  other uses will be sketched in the conclusion.
Interest in BNs in the healthcare community has increased over the last decade as for diagnosis and prognosis BNs represent a natural framework \textcolor{black}{ to analyze dependence among } risk factors. Furthermore, they can aggregate \textcolor{black}{ knowledge from experts, which is especially relevant in contexts in which data might be limited,} and still provide meaningful and accurate decision support \cite{mclachlan2020bayesian}.
Relevant work in the field includes \citet{wang2020survivability}, who propose a BN model for cancer treatment assessment and development monitoring; \citet{jang2024estimating}, who use a BN model together with expert knowledge to analyze the disease burden of breast cancer and the risks and benefits of radiation therapy; and \citet{liu2018quantitative} who use BNs to analyse the most influential factors in breast cancer diagnosis.
 Regarding CRC, \citet{myte2017untangling} build a BN to analyse the possible impact of one-carbon metabolites in relation to CRC, also considering genetic information and environmental factors in the study; \citet{sieswerda2023estimating} leverage BN structure learning algorithms and expert knowledge to create causal models to estimate treatment effectiveness in colon cancer therapies; and \citet{osong2022bayesian} make use of BNs for predicting local tumor recurrence in rectal cancer patients after treatment and surgery. 

In contrast, the approach proposed in this paper aims to build a representative \textcolor{black}{probabilistic} model of the interactions between several variables in a general population setting, including non-modifiable and modifiable risk factors, to analyze their influence in the development of CRC.
Major advantages of BN models, that we shall
draw upon, are their use for generative purposes and their ability to propagate the evidence along the network to obtain representative probabilities based on this evidence.
Thus, the model built in this paper is intended to serve as a quantitative guideline for the CRC risk assessment of different \textcolor{black}{segments of} a population, as it manages to maintain representative proportions and imbalances of the different variables found in the data set. Hence, the conclusions reached through the model will be representative (at least for the population set taken into account) and actions taken could be modeled to obtain a meaningful approximation of their \textcolor{black}{influence}. Furthermore, the characterization of segments of the population with a higher risk of developing CRC would be of interest for screening purposes as targeting these groups would yield more cases per screening test performed and increase a screening program's effectiveness \cite{ferlizza2021roadmap}. 

The rest of the paper is structured as follows. We first describe 
 how the BN was built taking into account
the data and knowledge available; this entails discovering the structure
of the network and building the corresponding tables of probabilities.
We next deal with two important use cases: the first one refers to building risk 
maps depending on key features of individuals; the second one, refers to 
 reporting key factors in developing CRC.
A final section summarises results, discusses limitations, suggests additional use cases, and 
sketches future work. 
\textcolor{black}{ Importantly, for reasons outlined in this last section,  we prevent from making causal claims for our BN and just 
pursue predictive claims as in \citet{hernan2023causal}; \citet{scutari2021bayesian} provide further insights regarding causality and BNs}.
For reproducibility purposes, software for the full model, 
as well as for the use cases presented, is available in \href{https://github.com/DanielCorralesAlonso/CRC_Risk_BN}{ https://github.com/DanielCorralesAlonso/CRC\_Risk\_BN}.

\section{Materials and methods}
\label{section2}
This section describes the process used to build our BN for CRC risk predictions. It is divided into five parts characterizing the work pipeline adopted: collection of available knowledge, data gathering and processing, network structure discovery, estimation of probabilities, and validation.


\subsection{Materials} 
\subsubsection{Prior available knowledge} \label{subsection2.1}

The data used in this project were extracted from an observational study covering 
   annual health assessments of adult workers affiliated with a private health insurance provider in Spain, from 2012 to 2016. After  
conveniently securitizing the data, they were enriched with census information
from the Spanish National Statistics Institute (INE) based on postal code, allowing us to infer
their socioeconomic status
and 
educational level. 
 This led to an initial dataset 
 with about 2.4 million records and 66 variables.

In order to compile relevant knowledge about CRC, we performed exhaustive searches through scientific and medical databases with the expressions
{\em `causal inference and CRC'; `probabilistic networks, Bayesian networks,
influence diagrams and CRC'; `Data mining and CRC'; `Risk factors and CRC'; `CRC epidemiology'; `causes of CRC'}. 
We also queried ChatGPT with the prompts
{\em 'What are the risk factors in the 
development of CRC'} and 
{\em 'What are the modifiable risk factors in the 
development of CRC'}.
Additionally, relevant information from a previous network developed 
 concerning cardiovascular disease (CVD) risk factors
\citep{CVD} was considered.

\subsubsection{Available data}\label{subsection2.2}

The list of relevant variables, together with the background
information mined, was submitted to a team of expert clinicians
who, through a consensus session, suggested 
 to retain from the original database the fourteen variables presented 
 in Table~\ref{tab:Variables} in the Appendix. 
  They also 
grouped  the variables as follows: 
\begin{itemize}
    \item Non-modifiable CRCRFs: {\em sex, age,} and {\em socioeconomic status}.
    \item Modifiable CRCRFs: {\em body mass index (BMI), \textcolor{black}{physical activity (PA)},
sleep duration (SD), alcohol consumption, smoking profile, anxiety,} and {\em depression}.
    \item Medical conditions: {\em hypertension, hypercholesterolemia,} and {\em diabetes}.
\item Target variable: {\em presence of CRC}.
\end{itemize}

\noindent An intensive exploratory data analysis focused on detecting 
outliers and misrecorded values, duplicates, and missing values.

In particular, 
\textcolor{black}{ for originally continuous unimodal approximately symmetrical variables around the mean, we considered the standard rule of treating as outliers those data points whose values were further from the marginal distribution's mean by three standard deviations \cite{wada2020outliers}, with 230,841 data points meeting these criteria.  These were assumed to come from }
measuring or recording mistakes; we removed them 
from the training phase, assuming that model performance would not be affected.\footnote{Importantly, 
 they were not used for training purposes, but we used them for
validation purposes in the sense of Section \ref{subsection2.5}}
  As an example, the case of a patient whose record showed that their height was 160cm and their weight 342kg,
  was eliminated. 
We also discarded for training purposes 325,147 data points with a missing value in any variable, as given the size of the final dataset, we would have enough training data. \textcolor{black}{Note that there was no evidence suggesting any missing not at random (MNAR) scenario which would have prevented us from discarding these data points.}\footnote{Again we used them for validation purposes, Section \ref{subsection2.5}} 

Finally, we retained a total of 1,778,270 health assessments which were split according to the date of the recording with the motivation of updating the parameters of our model every year based on information from previous years
 and reserving those of year 2016 for validation purposes.

 Table~\ref{tab:Empirical_Marginal_Distr} in the Appendix lists the proportion of cases in various 
  marginal categories reflecting, by and large, the standard structure of the Spanish labor force. We performed this exploratory analysis for each of the years as an exploratory sensitivity analysis check, \textcolor{black}{ revealing just minor differences over the years. }

    \textcolor{black}{Similarly, we explored the impact of spatial effects based on postcodes, finding no evidence of spatial correlations for all variables considered except the socioeconomic situation, in which spatial information is encoded by definition.}

\subsection{Methods}
\subsubsection{Structure discovery}\label{subsection2.3}

Once the data was collected and processed  we built a discrete BN to estimate the underlying joint distribution, which served as the basis to make inferences and predictions on CRC risk cases of interest.
The selected variables were coded as described by Table~\ref{tab:Variables} in the Appendix. A two-stage procedure
  was used to learn the BN structure.

\begin{figure}[t]
\centering
\includegraphics[width=1\textwidth]{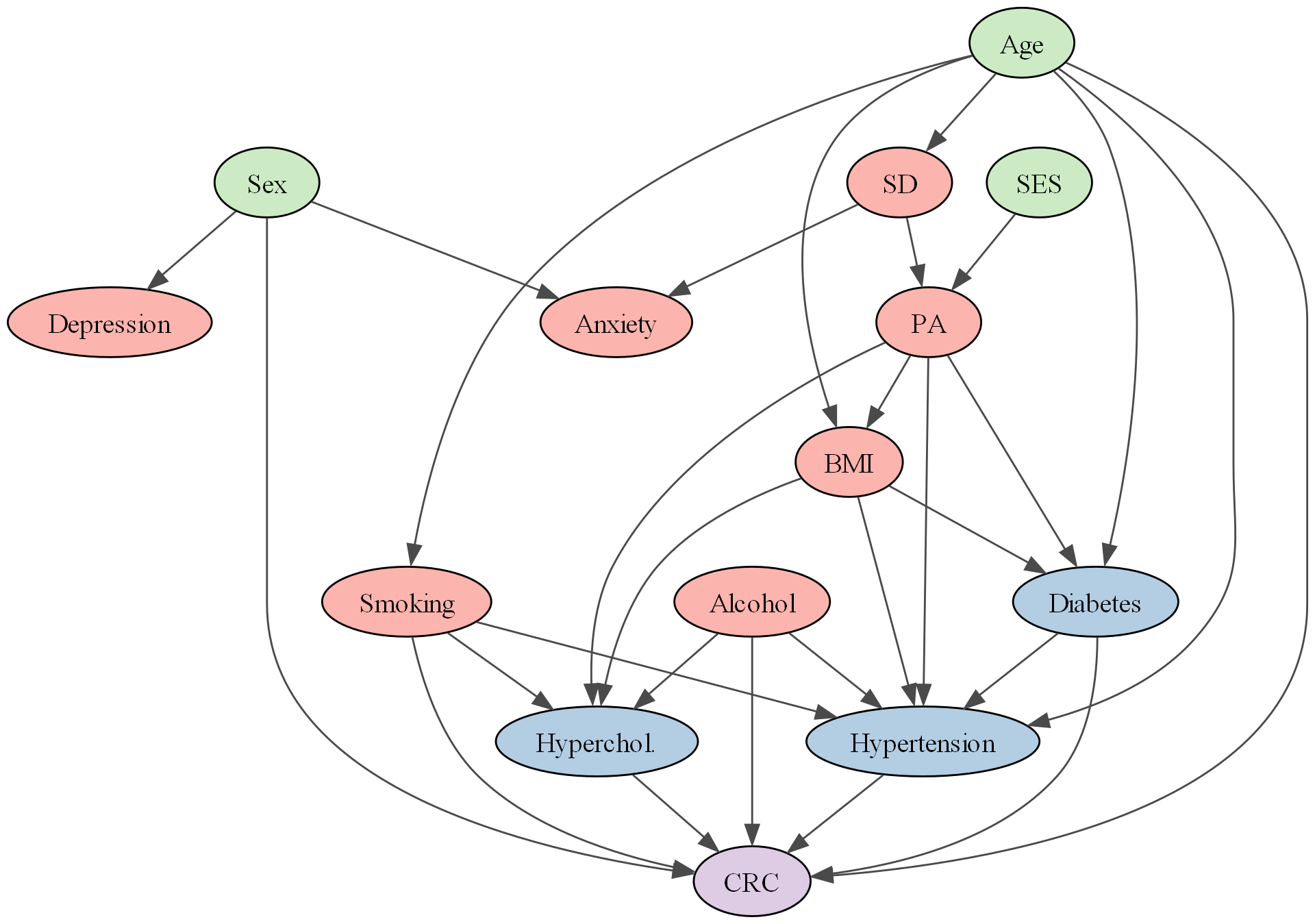}
\caption{Initial BN structure (network 1) coding knowledge available for CRC taking into account available variables. Forbidden arcs not included for clarity.}
 \label{fig:initial_final_layuout}

\end{figure}

 First, based on the information described in Sections \ref{subsection2.1} and \ref{subsection2.2}, specially 
 the causal suggestions from our medical experts, 
 \textcolor{black}{
 we obtained an initial description of the network describing proposed  and forbidden 
 arcs, summarising their knowledge, as agreed with the team of experts.  
Figure \ref{fig:initial_final_layuout} provides the   
initial network
  where, to facilitate visualization, we do not include forbidden arcs.}
 As an example, (Hypercholesterolemia, Age) would be a forbidden arc as the former cannot affect the latter in any possible way. Different color codes were used for the four types of
 variables mentioned above.

\begin{figure}[t]
\centering
\includegraphics[width=1\textwidth]{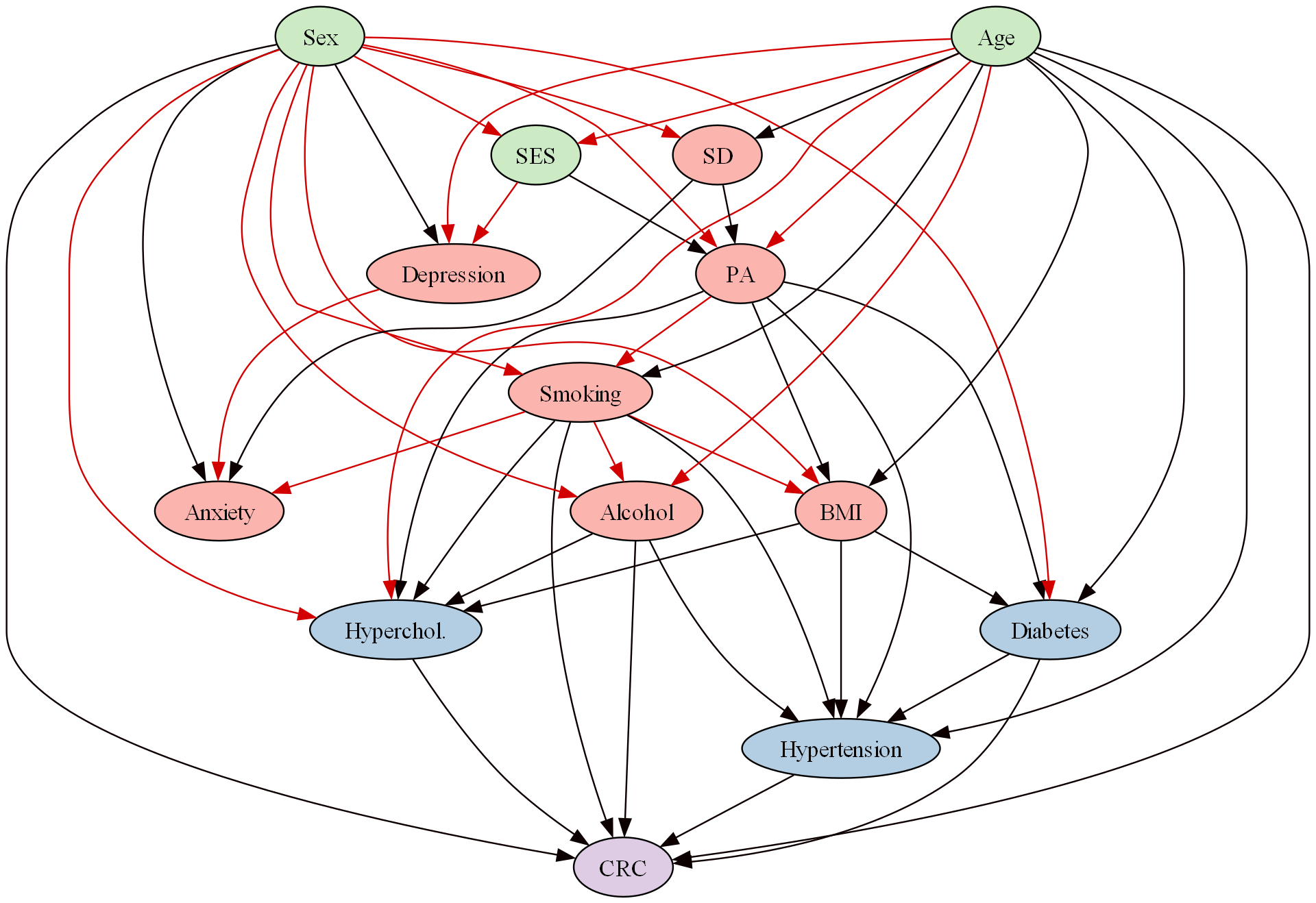}
\caption{Final BN structure (network 2) coding knowledge and data available for CRC taking into account relevant available variables and enhanced through the database. }
\label{fig:initial_final_layuout2}
\end{figure}

Such structure was used as the initial network to 
several structure discovery algorithms and software. 
There are numerous procedures available for the purpose of building a network based on relations in the data summarized e.g.\ in 
\cite{SCUTARI2019235} and \cite{scanagatta2019survey}, who also
 mention related software solutions. In particular, we used 
 the algorithms available in GeNIe 
 Modeler  \citep{bayesfusion2023genie}, 
 and the Python libraries
\emph{pyAgrum} \citep{hal-03135721} and \emph{pgmpy} \citep{ankan2015pgmpy}.
 The solutions arrived at with various algorithms were analyzed 
by three experts in the CRC domain who revised the additional arcs
reasoning in terms of plausible \textcolor{black}{predictive relationships.} 
This process led to 
the final BN structure shown in Figure~\ref{fig:initial_final_layuout2}
 where new data-based arrows are displayed in red. 
 \textcolor{black}{ To specifically obtain such a network, we employed the \textit{greedy hill-climbing} algorithm, a local optimization algorithm } that maximizes a predefined score at each step and adds an edge between nodes until the score cannot be maximized \cite{koller2009probabilistic}. \textcolor{black}{ For our network structure discovery,
   we used the Bayesian Dirichlet sparse (BDs) score defined in \citet{scutari2016empirical} and implemented in \emph{pgmpy}}\textcolor{black}{, as it is argued  \cite{scutari2018dirichlet} that BDs seems to provide better accuracy in structure learning, specially with sparse data.}
As a consequence of the chosen graphical
representation, the underlying suggested probabilistic model over the 
variables is characterized through the following expression:

{\small
\begin{align} \label{eq:1}
    p (v_{sex}, \dots ,v_{depression}) & =
    \Big[ p(v_{sex}) p(v_{age}) p(v_{SES}|v_{sex},v_{age}) \Big] \times \notag \\
  & \Big[  p(v_{SD}| v_{sex}, v_{age}) p(v_{PA}|v_{sex}, v_{age}, v_{SD}, v_{SES} ) p(v_{depr}|v_{sex}, v_{age}, v_{SES}) \notag \\
  &  p (v_{smok} | v_{sex}, v_{age}, v_{PA})
  p(v_{alc}|v_{sex}, v_{age}, v_{smok}) \notag \\
  & 
  p(v_{BMI}|v_{sex},v_{age}, v_{PA}, v_{smok})
  p(v_{anx} | v_{sex}, v_{SD}, v_{smok}, v_{depr}) \Big] \times \notag \\
    & \Big[ p(v_{hypchol}| v_{sex}, v_{age}, v_{PA}, v_{smok}, v_{BMI}, v_{alc})  p(v_{diab}| v_{sex}, v_{age}, v_{PA}, v_{BMI} ) \notag \\ 
    &
    p(v_{hypten}| v_{age}, v_{PA}, v_{smok}, v_{BMI}, v_{alc}, v_{diab})  \Big] \times  \notag
    \\
        & p(v_{CRC}|v_{sex}, v_{age}, v_{alc}, v_{smok}, v_{hypchol}, v_{hypten} , v_{diab} ) 
   \end{align}
}

\noindent where, to facilitate reading
and reasoning,  we have separated the products into the four blocks of variables considered.

\subsubsection{Probabilities discovery}\label{subsection2.4}

Once with the structure, the next stage was to learn the 
associated probability tables drawing on the data $D$ available.
We estimated them using standard multinomial-Dirichlet models 
 \citep{French,koller2009probabilistic}. \textcolor{black}{ Let $X$ be a network variable, $U$ its parent variables, and $\bm{u}$ one
  of its instantiations}. In general, if $p(\theta_{X | \bm{u}})$ is a Dirichlet prior distribution with hyperparameters $\alpha_{x^1 | \bm{u}},..., \alpha_{x^K | \bm{u}}$, the posterior $p(\theta_{X | \bm{u}} | D)$ will be 
  a Dirichlet distribution with hyperparameters $\alpha_{x^1 | \bm{u}} + m[\bm{u}, x^1],..., \alpha_{x^K | \bm{u}} + m[\bm{u}, x^K]$, 
  \textcolor{black}{ where $m[\bm{u},x^i]$ is the number of times that instance $(\bm{u},x^i)$ appears in the dataset}. In particular, the 
   estimate of the parameter $\theta_{X = x^i | \bm{u}}$ based on the posterior mean would be 
 \begin{equation*}
     \hat{\theta}_{X = x^i | \bm{u}} = \frac{\alpha_{x^i | \bm{u}} + m[\bm{u}, x^i]}{\sum_{i} \left( \alpha_{x^i | \bm{u}} + m[\bm{u}, x^i] \right)}.
 \end{equation*}
 
 A potential problem with our BN structure 
is that, due to the many connections arriving at some of the nodes some of the columns in the tables might \textcolor{black}{ receive relatively} little data. In particular, minority classes in highly imbalanced variables, e.g.\ the CRC positive class in the CRC node, are affected by this issue.
In that case, the corresponding posterior distributions would essentially coincide with the priors, \textcolor{black}{ therefore demanding care in assessing 
such priors,}
 notwithstanding the related problem of the large number of priors 
  to be chosen for some of the variables considered in the model.

 Uniform priors are largely used in scenarios where no prior information is available. \textcolor{black}{One example is the prior defined for the Bayesian Dirichlet equivalent uniform (BDeu) score \cite{buntine1991theory}, which assumes complete ignorance about the parameters of the network and thus at each node each class has the same probability \cite{castelo2000priors}, \cite{ueno2012learning}. In the case of the prior for the BDs score used for structure learning, it follows an empirical Bayes approach by giving prior uniform probability to the classes that appear at least once in the dataset, and zero prior probability to the classes that do not appear in the dataset \cite{scutari2016empirical}. Still}, in medical practice \textcolor{black}{the} lack of prior information is rare, and a carefully defined informative prior may be more \textcolor{black}{ meaningful} than a uniform one. 
As a consequence, the following approach was employed to build the priors \textcolor{black}{for estimating the parameters}. Table \ref{tab:Empirical_Marginal_Distr} in 
 Appendix \ref{appendix} provides the marginal empirical distributions for all variables,
 which we use as prior means for the corresponding conditionals, whatever the conditioning values are, \textcolor{black}{as a means of characterizing prior knowledge}. We multiply them by a factor $\alpha$ interpreted as the relative weight of the prior with respect to the data in the calculation of the posterior distribution. After cross-validating \cite{allen2000model}, \cite{scutari2019learning} this
parameter  by  trying several values in a grid, based on classification \textcolor{black}{performance} (section \ref{subsection2.5}) and \textcolor{black}{quality of} inference (sections \ref{section3} and \ref{section4}),
 $\alpha$ was set to \textcolor{black}{the number of patients considered divided by 10000, that is $\alpha \approx 31.69$
 ,} as it entailed a reasonable influence of the prior knowledge in the above-mentioned cases with few data when a variable is conditioned by many others. 
  \textcolor{black}{ Other $\alpha$ values were tried varying the denominator in powers of 10; some of their multiples resulted
  in poorer performance, in classification terms, in the extreme cases in which $\alpha$ was too small or too large; other 
   intermediate candidate values resulted in more similar performance to the $\alpha$ selected.}   \textcolor{black}{ A limitation of this approach is that the parameter $\alpha$  will have a different impact on the parameterization of each variable depending on its skewness or uniformity. This has been further analyzed in the literature, see \cite{ueno2011robust}, \cite{silander2007sensitivity}. However, it simplifies the prior characterization by only determining a single free parameter.}
  
  This quantity $\alpha$ will determine the uncertainty for all probability distributions at all nodes. 
  Then, as discussed, if there is sufficient data for each of the combinations of conditioning variables, the uncertainty for the distributions will be reduced and the posterior means will shift depending on the conditioning variables.
   For cases with less data, the posterior distributions will be more similar to each other but will entail a larger uncertainty of the approximation. This process is repeated over several years, from 2012 to 2015, using the posterior distribution of the previous year as the prior for the next one, appraising the value of data from previous years.
  \textcolor{black}{  As an example, Table  \ref{alphas} provides the prior means for the distribution of 
  the variable {\em SD} conditional on its two antecessors, {\em Age} and {\em Sex}, based on the marginals in
   Table \ref{tab:Empirical_Marginal_Distr} for each of the three categories Short (S), Normal (N), and Excessive (E), which,
   as described above, coincide.}

 \begin{table}[!ht]
 \footnotesize
 \centering
\begin{tabular}{|l|l|l|l|l|l|l|l|l|}
\cline{2-9}
       \multicolumn{1}{c|}{} & \multicolumn{2}{c|}{{[}24,34{]}} &    \multicolumn{2}{c|}{{[}34,44{]}} & 
       \multicolumn{2}{c|}{{[}44,54{]}} &    \multicolumn{2}{c|}{{[}54,64{]}}           \\ \cline{2-9}
       \multicolumn{1}{c|}{} & Man         & Woman      & Man         & Woman      &Man         & Woman      & Man         & Woman      \\ \hline
Short  & 0.1024   & 0.1024  & 0.1024  & 0.1024  & 0.1024   & 0.1024  & 0.1024   & 0.1024\\ \hline
Normal & 0.8963   & 0.8963  & 0.8963   & 0.8963  & 0.8963   & 0.8963  & 0.8963   & 0.8963\\ \hline
Excessive  & 0.0011  & 0.0011 & 0.0011  & 0.0011 & 0.0011  & 0.0011 & 0.0011  & 0.0011\\ \hline
\end{tabular}
\caption{Prior mean probability for SD given Sex and Age}
\label{alphas}
\end{table}

\noindent \textcolor{black}{ Tables \ref{alphas2} and \ref{alphas3} respectively provide the posterior conditional mean and 
     0.9 posterior predictive intervals after processing the data from  year 2012.
   Observe that there has been a reasonable change in the posterior conditional probability table when compared with the prior table, effectively addressing the differences among the states and the conditional states of the variables considered.}
  
\begin{table}[H]
\footnotesize
\centering
\begin{tabular}{|l|l|l|l|l|l|l|l|l|}
\cline{2-9}
       \multicolumn{1}{c|}{} & \multicolumn{2}{c|}{{[}24,34{]}} &    \multicolumn{2}{c|}{{[}34,44{]}} & 
       \multicolumn{2}{c|}{{[}44,54{]}} &    \multicolumn{2}{c|}{{[}54,64{]}}           \\ \cline{2-9}
       \multicolumn{1}{c|}{}& Man         & Woman      & Man         & Woman      & Man         & Woman      & Man         & Woman      \\ \hline
Short  & 0.0600  & 0.0711 & 0.0897  & 0.1039 & 0.1211  & 0.1581 & 0.1386  & 0.2256 \\ \hline
Normal & 0.9384   & 0.9264 & 0.9092  & 0.8952 & 0.8778  & 0.8407 & 0.8604  & 0.7737 \\ \hline
Excessive   & 0.0016  & 0.0025  & 0.0011  & 0.0009 & 0.0012  & 0.0012 & 0.0010   & 0.0007  \\ \hline
\end{tabular}
\caption{Posterior mean probability for SD given Sex and Age in 2012}
\label{alphas2}
\end{table}

\begin{table}[!ht]
    \tiny
    \centering
    \begin{tabular}{|>{\hspace{-1pt}}l<{\hspace{-1pt}}|>{\hspace{-1pt}}l<{\hspace{-1pt}}|>{\hspace{-1pt}}l<{\hspace{-1pt}}|>{\hspace{-1pt}}l<{\hspace{-1pt}}|>{\hspace{-1pt}}l<{\hspace{-1pt}}|>{\hspace{-1pt}}l<{\hspace{-1pt}}|>{\hspace{-1pt}}l<{\hspace{-1pt}}|>{\hspace{-1pt}}l<{\hspace{-1pt}}|>{\hspace{-1pt}}l<{\hspace{-1pt}}|}
    \cline{2-9}
       \multicolumn{1}{c|}{} & \multicolumn{2}{c|}{{[}24,34{]}} &    \multicolumn{2}{c|}{{[}34,44{]}} & 
       \multicolumn{2}{c|}{{[}44,54{]}} &    \multicolumn{2}{c|}{{[}54,64{]}}           \\ \cline{2-9}
       \multicolumn{1}{c|}{}& Man         & Woman      & Man         & Woman      & Man         & Woman      & Man         & Woman      \\ \hline
   S &    [.0583, .0617] & [.0686, .0737] & [.0881, .0914] & [.1013, .1064] & [.1189, .1232] & [.1543, .1619] & [.1347, .1425] & [.2175, .2338] \\ \hline
   N &     [.9367, .9401] & [.9238, .929] & [.9076, .9108] & [.8926, .8978] & [.8756, .88] & [.8369, .8445] & [.8565, .8643] & [.7654, .7818] \\ \hline
   E &  [.0013, .0019] & [.002, .003] & [.0009, .0013] & [.0007, .0012] & [.0009, .0014] & [.0008, .0015] & [.0007, .0014] & [.0003, .0013] \\ \hline

    \end{tabular}
\caption{ 0.9 posterior predictive interval for SD probability given Sex and Age after processing 2012 data.}
\label{alphas3}
\end{table}

\noindent 
\textcolor{black}{ \noindent Table \ref{alphas_full} provides the evolution of the SD probability distribution for a man in the age range $[24,34]$ after processing the data from years 2012 to 2015,
  displaying the mean and the 0.9 posterior predictive intervals for each year. The aforementioned change in the posterior probabilities is more subtle after 2012 as the prior information of the previous years was already informative. 
  Nevertheless, this approach is able to detect subtle changes in the distribution over the years which can be highly useful in certain contexts.}

\begin{table}[H]
    \scriptsize
    \centering
    \begin{tabular}{|l|l|l|l|l|l|l|l|l|l|}
    \cline{2-10}
        \multicolumn{1}{c|}{} & Prior & \multicolumn{2}{c|}{2012} &  \multicolumn{2}{c|}{2013} &  \multicolumn{2}{c|}{2014} &  \multicolumn{2}{c|}{2015}  \\ \hline
        Short & .1024 & .0600 & [.0583, .0617] & .0600 & [.0581, .0613] & .0595 & [.0579, .0612] & .0608 & [.0591, .0626] \\ \hline
        Normal & .8963 & .9384 & [.9367, .9401] & .9388 & [.9372, .9404] & .9389 & [.9373, .9406] & .9378 & [.9360, .9396] \\ \hline
        Excessive & .0011 & .0016 & [.0013, .0019] & .0015 & [.0013, .0018] & .0015 & [.0013, .0018] & .0013 & [.0011, .0016] \\ \hline
    \end{tabular}

\caption{Evolution of the SD probability distribution for a man in the age range [24-34] 
 over years 2012-2015.}
\label{alphas_full}
\end{table}

The proposed method seeks to address the aforementioned problem related to the priors through the implementation of informative and representative priors, which in the cases where none or few data are collected avoids assessing uniform probabilities to combinations of variables that are so rare that might not even appear in the data.
A uniform probability prior in this scenario would represent exactly the opposite of what we infer from the data as would characterize these combinations of variables with around a probability of \textcolor{black}{$1/k$ (for $k$-valued categorical variables)} in the conditional distributions of the model, resulting in a poor and misleading probability assessment.
 By acknowledging these situations, we reduce the uncertainty surrounding the less frequent values and we shall better characterize the risk assessments to be performed.

\subsubsection{Validation through classification}\label{subsection2.5}

 Once the model has been built, and before illustrating relevant use cases 
 in Sections \ref{section3} and \ref{section4}, a core issue is to validate it. 
    A natural way to do it \textcolor{black}{in a probabilistic setting,
     see e.g. \cite{bielza2014discrete}, \cite{allen2000model} and \cite{scutari2019learning}},
    is to conceive the network as a classifier and assess its performance over various nodes with a number of classification metrics. 
    We undertook extensively this approach suggesting good results. 
    
   \textcolor{black}{ Let us illustrate the process with two variables, CRC and Diabetes.} 
  In the first case, we set CRC as the target variable that we would like to classify using the 
  available instances for 2016, and those related to outliers and missing values.
  Note that the problem we are dealing with in this case is a highly imbalanced problem (1:1500 approx) 
    which entails a major challenge for classifiers \citep{imbalanced}. 
   As an example, using the BN built, we classify the data set for the 2016 patients and 
    aim to maximize the {\em G-mean}, the root of the product between sensitivity and specificity \citep{gmean}. Recall that the major interest will be to detect as many CRC positives as possible without falsely classifying CRC negatives as positives. Table \ref{confmatCRC} 
    presents the confusion matrix achieved in the classification 
    of the whole data set, achieving a sensitivity of 0.68 and a specificity of 0.72. The corresponding AUC score is 0.76, which, incidentally, surpasses the values reported by other CRC studies with similar datasets, population imbalance characteristics, and calibration results \cite{smith2018added}.
\textcolor{black}{ In the case of diabetes, the classification problem is 
much less imbalanced (1:30). Table \ref{confmatDiab} provides the confusion matrix 
achieved, with a sensitivity of 0.73 and a specificity of 0.76.}

\begin{table}[h]
    \begin{subtable}[h]{0.5\textwidth}
        \centering
        \begin{tabular}{|c|c|c|c|}
        \cline{3-4}
           \multicolumn{2}{c|}{} & \multicolumn{2}{c|}{Pred. label}  \\ \cline{3-4}
           \multicolumn{2}{c|}{} & 0 & 1 \\ \hline
            \multirow{2}{*}{True label}         & 0     & $\bm{243326}$        & 96163      \\ \cline{2-4}
                   & 1    & 70        & $\bm{148}$  \\ \hline  
    
        \end{tabular}
        \caption{CRC Confusion Table}
        \label{confmatCRC}
    \end{subtable}
    \hfill
    \begin{subtable}[h]{0.5\textwidth}
        \centering
        \begin{tabular}{|c|c|c|c|}
        \cline{3-4}
           \multicolumn{2}{c|}{} & \multicolumn{2}{c|}{Pred. label}  \\ \cline{3-4}
           \multicolumn{2}{c|}{} & 0 & 1 \\ \hline
                \multirow{2}{*}{True label}       & 0     & $\bm{249937}$        & 78361      \\ \cline{2-4}
                       & 1    & 3118       & $\bm{8291}$  \\ \hline  
        
        \end{tabular}
        \caption{Diabetes Confusion Table}
        \label{confmatDiab}
    \end{subtable}
    \caption{Confusion Tables for BN validation}
\end{table}

\begin{figure}[h]
\centering
\includegraphics[width=0.3\textwidth]{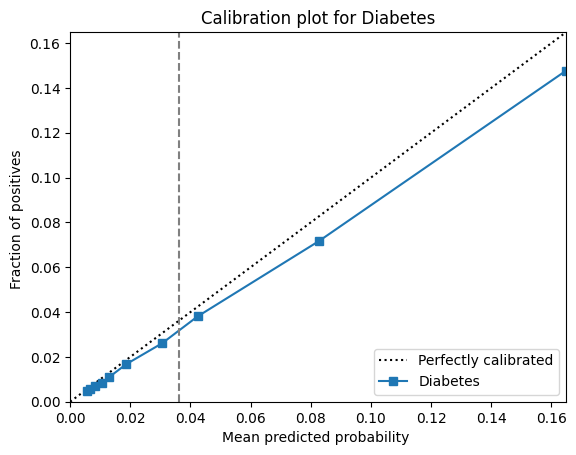}
\includegraphics[width=0.3\textwidth]{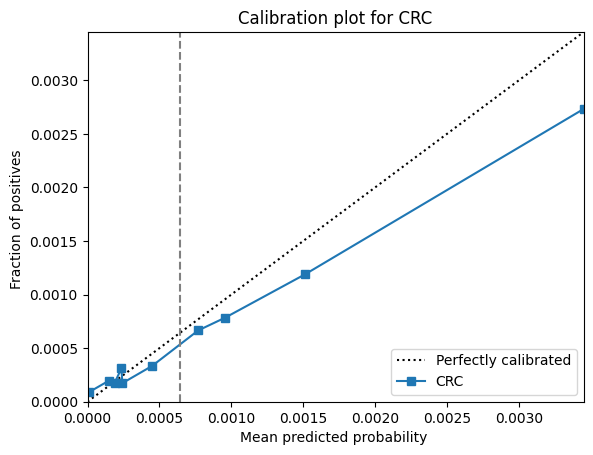}
\caption{Calibration curves for Diabetes and CRC }
 \label{calibration}
\end{figure}

 Besides the usual classification metrics, we paid special attention to their 
 calibration, in line with recent discussions in the medical literature \citep{van2019calibration}.
  This is of vital importance in risk prediction models as it has a great impact on the 
  usefulness of these decision-support aspects. Figure \ref{calibration} displays the calibration curves obtained through quantile binning for the cases of Diabetes and CRC over the relevant ranges for both diseases.\footnote{The empirical marginal of diabetes in 2016 is 0.0336 and that of  CRC is 0.00064.} 
  Quantile binning \cite{naeini2016binary} creates bins with an equal number of samples based on the distribution of the data instead of bins with equal width. Thus, the number of predictions is larger on the lower end of the distribution in the cases of imbalanced data and fewer predictions are made on the upper end of the distribution. The resulting curves suggest a good calibration with a slight tendency to overestimate in the final relevant bins.

\section{Results}
\subsection{Use case: CRC risk mapping}\label{section3}

Once the BN has been built, parameterized, and validated, we proceed to exploit some of its properties and functionalities. The first use case for our model is the production of risk maps or tables that reflect the risk of a person suffering CRC assuming certain conditions $c$ (e.g., this person is a man who is a smoker) as other features $b$ vary (e.g., his age and drinking status) of interest.  The motivation behind this use case is the well-evidenced assumption that different conditions have non-identical CRC effects in distinct segments of the population \cite{Marley2016-hl}. Furthermore, eventual tendencies could be broadly characterized through the use of risk maps.

The basic ingredient for the design of this tool would be the probabilities $p(CRC|c,b,q)$ of a person having CRC given that it has features $b$  and $c$, as $b$ adopts values in a set $B$, when $q$ are the parameter values adopted for the probability tables, which are computed from the BN model with standard Bayesian computations \cite{French}.To facilitate interpretation, we perform a comparison against the baseline of not having the
  information $b$, computing the differences in log probabilities 
\[  r(b,q)= \log (p(CRC|c,b,q))- \log (p(CRC|c,q)) , \]
and display graphically such quantities
as a function of $b$.  Recall though that we have uncertainty about $q$ and thus we have 
to reflect it, for example through an interval $i(b)= \left[ lr (b,q), ur (b,q) \right]$
 of high posterior predictive probability  for $r(b,q)$. For that, an iterative sampling approach is followed to generate 
 \textcolor{black}{ posterior predictive estimates for the probabilities }  of interest. The uncertainty is then
 reflected through the, \textcolor{black}{ e.g.,} 0.9 posterior predictive interval of the desired quantity
 and, essentially, we would declare that if:
 \begin{itemize}
 \item    $0 \in i(b)$ there is no sufficient evidence for an increase in risk
  with respect to the baseline; 
  \item  $0< lr (b,q) $, there is an increase in risk; and, finally,
  \item   $ 0 >  ur (b,q) $  there is a reduction in risk.
  \end{itemize}
  
After several design and visualization tests, we decided to display the risk maps as follows:
\begin{itemize}
\item Condition $b$ would refer to one or two criteria, leading to uni- or bi-dimensional risk 
maps.
\item We use $r(b,\hat{q})$ as reference for graphical purposes,  where $\hat{q}$ is the posterior mean of $q$,
 \textcolor{black}{  but additionally include} $i(b)$.
\item  \textcolor{black}{ A color scheme based on $r(b,\hat{q})$ is used and} displayed together with the whole interval $i(b)$.
We avoid colors typically used in risk matrices \citep{anthony2008s} (red, yellow, green) to mitigate cultural biases.
\item The size of the representation associated with the variation of risk in the segment $b$ 
should reflect the size of the corresponding population.
\end{itemize}

\noindent We provide now several examples of risk maps based on the previous guidelines.

\begin{figure}[H]
\centering
\includegraphics[width=0.6\textwidth]{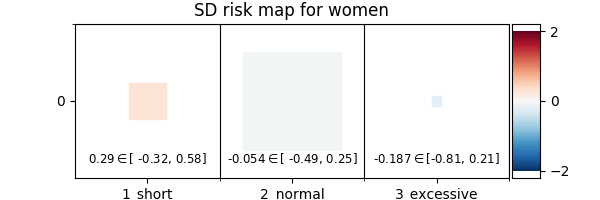}
\caption{Risk map for {\em sleep duration (SD)}  for $women$ }
 \label{risk:map3}
\end{figure}

 \paragraph*{Example 1. }
The first example, Figure \ref{risk:map3}, provides a risk  map when  $c=$ {\em woman}, taking into account 
$b=$ ({\em SD}) reflected in the $x$-axis, that is, we want to display the CRC risk variation depending on the sleep duration (short, normal, excessive) in women. \textcolor{black}{ Therefore, the} reference probabilities are
$p(CRC|woman,SD,q)$. 

 In this case, shorter sleep duration seems to be related to an 
 increase in CRC risk as shown by the point-wise estimations reflected in the colors and the first quantity in each of the cells.
  However, the interval
 estimates do not confirm this finding as $0$ belongs to all the 0.9 posterior predictive intervals. Therefore, we would conclude that SD is not a variable that fundamentally increases the risk of CRC on its own.
 Observe that the normal SD group is the largest one, followed by a smaller group with shorter SD. Note also that the smaller the population group, the larger the uncertainty as shown by the lower and upper bounds of the reported 0.9 posterior predictive intervals.\hfill $\triangle$
 
\begin{figure}[H]
\centering
    \includegraphics[width=0.7\textwidth]{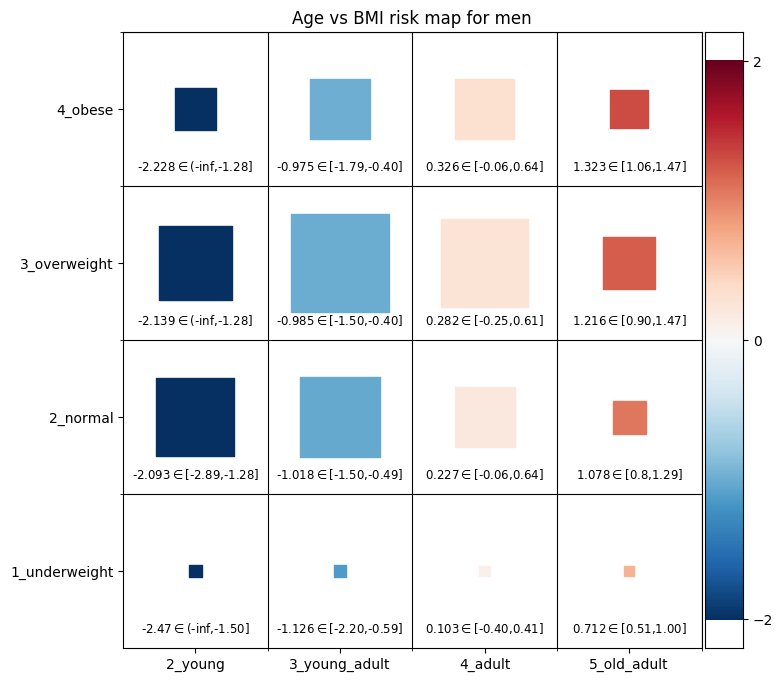}
    \caption{Risk map for $Age$ and $BMI$ for $men$ }
\label{risk:map1}
\end{figure}

\paragraph*{Example 2.}
Figure \ref{risk:map1} provides a risk  map when $c= man$,  taking into account  that $b=(Age, BMI)$
with $age$ varying in the $x$-axis and $BMI$ in the $y$-axis. Thus, the reference probabilities
are $p(CRC|man, (Age,BMI), q)$.

  Observe that CRC risk increases as both BMI and Age increase. However, age is the variable that has a larger impact, as colors are more similar column- than row-wise. We state that there is a smaller risk of CRC development with respect to the baseline for patients with ages lower than 44 and 
a bigger risk for patients older than 54.

\begin{figure}[ht]
\centering
\includegraphics[width=0.7\textwidth]{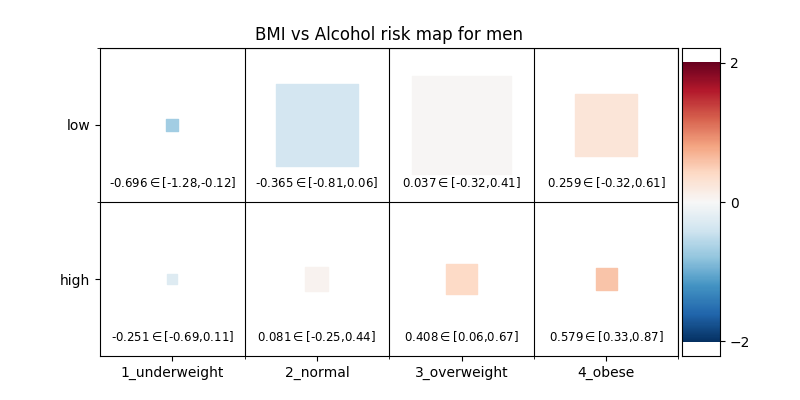}
\caption{Risk map for $BMI$ and $alcohol$ for $men$ }
 \label{risk:map2}
\end{figure}

In turn, Figure \ref{risk:map2} provides a risk  map for $c=man$ taking into account 
$b=(BMI, Alcohol)$ with $BMI$ in $x$-axis and $alcohol$ in $y$-axis, with  
reference probabilities defined through $p(CRC|man,(BMI, Alcohol), q)$.

 \noindent In this case, higher alcohol consumption always induces an increased 
 CRC risk which accentuates greatly with age. Moreover, alcohol consumption seems to influence CRC risk more than BMI.  \hfill $\triangle$

\subsection{Use case: influential findings}\label{section4}

\textcolor{black}{ Risk maps provide visual comparisons 
of population groups in terms of different risk factors.
An additional useful approach to the analysis of the factors potentially affecting the development of CRC would be to examine the variables that had the largest impact on patients diagnosed with CRC. 
 In line with Section 3.1 and earlier work in determining influential
 findings in BNs, e.g. \cite{laurit},  we propose an approach to 
  characterize the predictive power of each class and variable in the network. In our analysis, the variables will be modified independently among all the possible values for each risk factor and the difference in risk will be assessed. Repeating this with all CRC-positive patients in the database, we obtain an estimation of the strength of the predictive influence for each of the risk factors.
 As mentioned in the introduction, it is important, though, to
   remark that the influence of the variables depends on the model's graphical 
 structure, and any causality claim should be carefully analyzed before taking it for granted,
 see our final discussion. This prevents us from employing standard causal evaluations of effect sizes through interventions/do-calculus 
 or counterfactuals.}

 In detail, we proceed as follows, where Algorithm \ref{alg:python-to-pseudo} 
 summarizes the method used.
  First, the entire information of each CRC-positive patient
is recovered from the database. 
The order of the evidence available for a patient is randomized and set variable by variable. At each step,  the relative
risk variation is calculated, which is quantified as the relative change in the difference of logarithms of the mean probabilities of developing CRC conditioned on the added evidence, similarly to the approach in Section \ref{section3}. That is,

\begin{equation*}
    RRV(i,j) = \frac{\log(p_{model}(CRC| ev_j) - log(p_{model}(CRC | ev_{j-1}))}{log(p_{model}(CRC | ev_{j-1})} \times 100 ,
\end{equation*}
   
\noindent where $RRV(i,j)$ refers to the relative risk variation for patient $i$ and variable $j$, and $ev_j$ represents the values of the first $j$ conditioning variables.

The reason for randomizing the evidence is that, when the evidence of the parents of the target node is fully set, the remaining variables have no effect on the target node as the entire probability distribution is determined by the parents of such node, due to the local Markov property \citep{koller2009probabilistic}. Thus, the order in which the evidence is set may have an impact on how certain variables seem to \textcolor{black}{influence the prediction on} the model target. 
Recording the relative variations in probability corresponding to the set of new evidence for each variable will assess the relative impact of the variable instance in the determination of the final probability.  Randomizing the order of the evidence and repeating the process several times 
  would provide a better understanding of the \textcolor{black}{predictive influence} of all the variables on the target node.

\RestyleAlgo{ruled}

\begin{algorithm}
\caption{Pseudo code to determine influential findings}
\label{alg:python-to-pseudo}

\KwData{Dataset, model, target}
\KwResult{diff\_vect}
\For{n iterations}{
    \For{row \textbf{in} rowsDataset}{
        \emph{evidence} = Dataset[row,:] \#Take variable information as evidence.

        $p_{model}(target | evidence)$ 

        \emph{shuffled\_evid} = random.shuffle(\emph{evidence})
        
        \For{$j \leftarrow 1$ \KwTo  \emph{len(shuffled\_evid)}}{
            \emph{partial\_evid} = Dataset[row, shuffled\_evid[0:j-1]]
            
            \emph{new\_evid} = Dataset[row, shuffled\_evid[j]]

            \emph{relative\_risk\_variation[row, j]} = $\frac{\log(p_{model}(target | partial\_evid + new\_evid)) - log(p_{model}(target | partial\_evid))}{log(p_{model}(target | partial\_evid))} \times 100$
 
        } 
    }
    Average along the data set rows
}
Average along all iterations
\end{algorithm}

\begin{figure}[htb]
\centering
\includegraphics[width=0.45\textwidth]{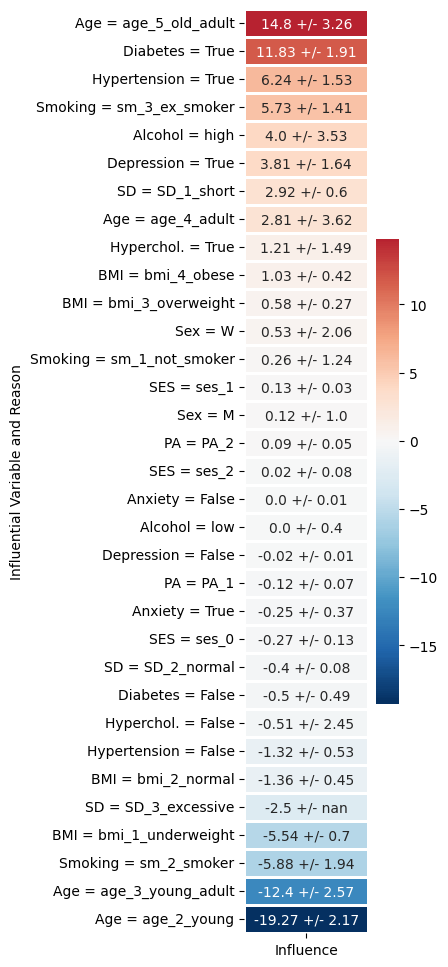}
\caption{Ranking of influential variables}
 \label{fig:Influence mapping std}
 \end{figure}

Figure \ref{fig:Influence mapping std} reflects an average of the positive and negative \textcolor{black}{predictive} influence that different variables have on the risk of developing CRC. \textcolor{black}{The standard deviations of the predictions are also provided}. Our conclusions seem to agree with \citet{GBD_2019_Colorectal_Cancer_Collaborators2022-eh} and \citet{Marley2016-hl},
which state that countries in Western Europe are prone to an increased consumption of alcohol and tobacco that highly contributes to CRC DALYs (Disability Adjusted Life Years). Furthermore, high fasting plasma glucose is one of the major contributors to CRC DALYs in Western European women and our analysis coincides with this by showing how diabetes is one of the main influential factors in the development of CRC.
Although not modifiable, age is certainly the most significant factor influencing the risk of developing CRC as about 90\% of the new cases occur in individuals over 50 years old \citep{sawicki2021}. Moreover, a larger BMI seems to affect also the risk of developing CRC.

The influence of smoking in our model is interesting as it would seem that it is better to be a smoker than to quit tobacco and become an ex-smoker. This appears to be related to the fact that the effects of smoking on CRC are mainly observed in the long run. People tend to be smokers when they are young and quit tobacco when they become older or are diagnosed with some condition for which tobacco is known to be a risk factor. \textcolor{black}{Furthermore, as we are in the context of an observational study, we cannot discard the possibility that heavy smokers may have died earlier due to other conditions not recorded in the study}. Thus, it is being an ex-smoker that would determine the risk of smoking in this case. However, further analysis would have to be done to reach a definitive conclusion.

Similar studies could also be performed using just certain segments
of the CRC-positive population, which could target more precisely the influence of relevant factors in a specific group.

\section{Discussion}

The proposed BN associates relevant medical conditions and CRCRFs 
in relation to CRC.
We used expert opinion to get its initial structure and an 
 extensive database \textcolor{black}{ to update and complement it}, from which 
 we also built its conditional probability tables, 
 with \textcolor{black}{ uncertainty in the beliefs acknowledged} through posterior distributions.
 
We illustrated its use to provide risk maps and uncover CRC influential variables.
But there are other relevant medical use cases
\textcolor{black}{  which we briefly sketch:
\begin{itemize}
\item As mentioned, we had access to individuals' postcodes.
This enables displaying geographical risk maps similar to those 
of section 3.1 with the 
whole country as baseline and
cells representing, say, provinces and their population size.
\item  Another important use is 
the classification of individuals, which 
we sketched in Section \ref{subsection2.5} for validation purposes, facilitating 
classifying an individual as more likely than not to have CRC.  
  Should a different  
utility function be available, we would assign individuals to the class with maximum 
expected utility. 
\item In turn, and similarly, we could use the BN to segment a population 
based on posterior CRC probabilities or posterior expected utilities, given certain features, say for screening purposes, as we shall 
do in future work.
\item A further important application of the network is for synthetic data generation purposes
when available data are proprietary and we need to share the 
data with a related organization \cite{kaur2021application}; this is easily achieved by sampling 
from the model defined in (\ref{eq:1}).
\item A collateral use of our BN would be to generate 
interesting medical hypothesis. As an example, Tables \ref{alphas2} and \ref{alphas3} show how sleep duration is affected by age, as older people seem to sleep for shorter periods than younger people. There also seems to be a significant gap between men and women in terms of sleep duration being women the ones that sleep less, with this gap accentuated with age.
\end{itemize}}

\textcolor{black}{ Our discussion in Section \ref{subsection2.4} about the prior chosen} reflected the important dynamical aspect of updating the initial prior through the data over various years. \textcolor{black}{ This is of interest as the model can be easily updated to consider the most recent data acquired by the health insurance provider in order to be used again for risk assessment purposes with up-to-date information.}

In future work, we shall
incorporate this predictive model into the larger decision-support picture
related to coherently advising screening methods. 
For this, we would need to consider the possible overall impact of the medical conditions using decision variables and utility functions. A decision-making problem will be defined for which the goal would be to find the portfolio of 
 screening recommendations with maximum expected utility in 
 line with \textcolor{black}{  precision vs current one-size-fits-all based on age }  approaches to screening \citep{KASTRINOS}.
   \textcolor{black}{ Such model would facilitate the design } of incentives to promote 
the adoption of CRC screening mechanisms and overcome current low adoption rates.

 We conclude by pointing out several limitations of this study.
 First, the exploratory analysis described in Table \ref{tab:Empirical_Marginal_Distr} suggests 
  a labor structure most probably different to that in other countries meaning that this model would either have to be adapted to the population structure in those countries or be used with some care taking into account this fact; \textcolor{black}{ yet the broad pipeline described would
  be reproducible}.
Second, some of the data were self-reported; \textcolor{black}{ 
  however any possible fault was mitigated
by several quality control strategies as described in \cite{Ley}}. Third, we had no data available concerning diet, genetics, and gut microbiome data;
 BMI, diabetes, and hypercholesterolemia might partly account for diet  information, but this would be a confounding variable;
 concerning genetics, \citet{Marley2016-hl} claim that 
  about 35$\%$ of the CRC development risk is due to genes
  positively or negatively
  influencing patients. \textcolor{black}{ Very importantly, as mentioned above,}
  the absence of the above three factors would prevent from causality claims in this study. Note though, \textcolor{black}{ again as discussed above,
  that we could 
  anyway conclude predictive claims in the sense of } 
  \citet{hernan2023causal}, much as we did above in relation to sleeping duration.  Finally, also hinted above, although we have updated the model over the years, it would also be of interest to consider the case of a dynamic BN framework to model disease evolution over time. This approach would aid also 
  in extricating some cause-effect relationships between the variables.


\subsection*{Credit authorship contribution statement}
\textbf{D. Corrales}: Conceptualization, Methodology, Software, Formal analysis, Writing - Original Draft, Writing - Review and Editing;
\textbf{A. Santos}: Validation, Resources, Writing - Review and Editing;
\textbf{S. Lopez}: Validation, Data Curation, Writing - Review and Editing;
\textbf{A. Lucia}: Validation, Resources, Writing - Review and Editing;
\textbf{D. Rios Insua}: Conceptualization, Formal analysis, Writing - Original Draft, Supervision, Funding acquisition, Writing - Review and Editing

\subsection*{Funding}
This work was supported by the AXA-ICMAT Chair in Adversarial Risk Analysis; the Spanish Ministry of Science project 
PID2021-124662OB-I00;  and the European Union's Horizon 2020 Research and Innovation Programme under Grant Agreement N. 101097036 (ONCOSCREEN).

\subsection*{Declaration of competing interest}
We confirm that there are no conflicts of interest associated with this publication.

\subsection*{Acknowledgments}
We are grateful to Quirónprevención for the provision of data. Discussions with Victoria 
Ley, Victor Lopez, and Isabela Rios were very useful.

\appendix
\section{Data used}\label{appendix}

Table \ref{tab:Variables} provides the states  
of the fourteen variables used and how they are coded.

\begin{table}[htbp]
    \centering
    \footnotesize
    \begin{tabular}{cccc}
       {\bf Variable}    & {\bf Definition}  & {\bf Levels}  
       \\ [1ex] \hline 
       $ v_{sex} $   & Sex  & \{\emph{female, male}\}   \\ 
       
       $ v_{age} $   & Age  & (24\textcolor{black}{,}34], (34\textcolor{black}{,}44], (44\textcolor{black}{,}54], (54\textcolor{black}{,}64]   \\
       
       $ v_{SES} $   & Socioeconomic status  & \{1,2,3\}  \\
       \hline
       $ v_{BMI} $   & Body mass index  & \{\emph{underw., normal, overw., obese}\}   \\
       
      $ v_{PA} $   & Physical activity  & \{\emph{insufficiently active (1), sufficiently active (2)}\}     \\
      
      $ v_{SD} $   & Sleep duration  & \{\emph{short, normal, excessive\}} \\

       $ v_{alc} $   & Alcohol consumption  & \{\emph{low, high}\}    \\
       
       $ v_{smok} $   & Smoker profile  & \{\emph{non-smoker, ex-smoker, smoker}\}    \\ 
       $ v_{anx} $   & Anxiety  & \{\emph{yes, no}\}    \\
            $ v_{dep} $   & Depression & \{\emph{yes, no}\}   \\
       \hline
        $ v_{hypten} $   & Hypertension  & \{\emph{yes, no}\}  \\
       
        $ v_{hypchol} $   & Hypercholesterolemia  & \{\emph{yes, no}\}  \\
       
       $ v_{diab} $   & Diabetes & \{\emph{yes, no}\}    \\
        \hline
       $ v_{CRC} $   & Colorectal cancer & \{\emph{yes, no}\}    \\
      
       
    \end{tabular}
    \caption{Fourteen variables in the model\textcolor{black}{.}}
    \label{tab:Variables}
\end{table}
We briefly discuss how key variables were categorized. 
Age was divided into four groups
((24,34], (34,44], (44,54], and (54,64]), using the INE National Sport Habits survey coding, as in \cite{Ley}. The socioeconomic status, originally a continuous variable, was discretized 
 in three levels by binning its values using specified quantiles based on the variable's mean and standard deviation, with a larger index indicating a higher socioeconomic level. 
 
 Concerning BMI, we used the four WHO classes: {\em underweight} ($<$ 18.5 kg/m$^2$), {\em normal weight} ([18.5, 25) kg/m$^2$), {\em overweight} ([25, 30) kg/m$^2$),  and {\em obese} ( $\geq$ 30 kg/m$^2$). Participants' leisure-time PA levels were assessed as in \cite{CVD}, distinguishing between patients not meeting WHO minimum recommendations for aerobic PA in adults ({\em insufficiently active})
 and meeting them ({\em regularly active}).  SD was categorized as {\em short} (less than 6 hours), {\em normal} (6-9 hours), and {\em excessive} ($>$ 9 hours). The smoker profile reflected whether the patient was
 an active smoker, had never smoked, or was an ex-smoker.
 We also extracted whether the patient had anxiety or depression.
 
  Concerning medical conditions,  we used the following criteria:
  {\em  diabetes},  medicated for it or glycemia $\geq 125$ mg/dL;  {\em hypercholesterolemia}, medicated for it or  LDL $\geq 130 mg/dL$, HDL $\leq 40 mg/dL$, triglycerides $\geq 150 mg/dL$ or total cholesterol $\geq 200$ mg/dL; {\em hypertension}, medicated for it or systolic/diastolic blood pressure $\geq 139$/$90$mm Hg.

Table \ref{tab:Empirical_Marginal_Distr} describes the full dataset distribution over 
  all the years.
\small
\begin{table}[!htb]
   \centering
   \footnotesize
   \begin{tabularx}{\textwidth}{lllp{0.01cm}lll}
      \textbf{Variable} & \textbf{States} & \textbf{Marginal} & & \textbf{Variable} & \textbf{States} & \textbf{Marginal}
      \\ [1ex] \hline \\ [-1ex]

      Sex  & Female & 30.68 \% & & Physical Act. & 1 & 47.21 \% \\
        & Male & 69.32 \% & &  & 2 & 52.79 \% \\
        & & & & & & \\

      Age(y) & \textcolor{black}{(24,34]} & 21.21 \% & & Anxiety & Yes & 2.70 \%\\
       & \textcolor{black}{(34,44]} & 38.02 \% & & & No & 97.30 \% \\
       & \textcolor{black}{(44,54]} & 29.03 \% & & & &\\
       & \textcolor{black}{(54,64]} & 11.73 \% & & Sleep Dur. &  $<$ 6h & 10.88 \%  \\ 
       & & & & & (6h-9h) & 89.01 \% \\

      Socioeconomic  & 1 & 23.93 \% & & & $>$ 9h & 0.11 \%   \\
       status & 2 & 61.97 \% & & & &  \\
       & 3 & 14.10 \% & & Depression & Yes & 0.47 \%   \\

      & &  & & & No  & 99.53 \%  \\
       BMI & Underweight & 1.10 \% & & & &   \\
       & Normal & 41.27 \%& & Diabetes & Yes & 3.63 \% \\
       & Overweight & 40.67 \%  & & & No & 96.37 \%   \\  
       & Obese & 16.96 \% & & & &   \\
      & & & & Hypertension & Yes & 15.05 \% \\
      Smoker  & Non-Smoker & 49.90 \% & & & No & 84.95 \%   \\
      profile & Ex-Smoker & 30.16 \% & & & & \\
      
      & Smoker & 19.94 \%  & & Hypercholest. & Yes & 51.32 \% \\
       & &  &  &   & No & 48.68 \% \\
      Alcohol & low & 95.05 \%  & &  & & \\
      & high & 4.95 \%  & & CRC & Yes & 0.07\% \\
      & & & & & No & 99.93 \%   \\
   \end{tabularx}
   \caption{Percentage of observations at each class for variables in the model.}
   \label{tab:Empirical_Marginal_Distr}
\end{table}
\normalsize
With the exception of the lower presence of females, due to the 
 labor sectors served by the incumbent health insurance provider, the 
  structure and its health status seem by and large representative of the Spanish labor market. A {\em healthy worker effect} \cite{brown2017healthy} might explain  some of the somewhat lower estimates (anxiety, depression,
  diabetes).


\newpage
\bibliography{sample}

\end{document}